# Incorporation and control of defects with quantum functionality during sublimation growth of cubic silicon carbide

Michael Schöler, Maximilian W. Lederer, Philipp Schuh, and Peter J. Wellmann*

Friedrich-Alexander University Erlangen-Nürnberg (FAU), Materials Department 6 (i-MEET), Crystal Growth Lab, Erlangen, 91058, Germany
* Corresponding author: peter.wellmann@fau.de

Superconductor based quantum computing has the major drawback of working temperatures which require liquid helium for cooling. A promising approach to overcome this obstacle for quantum technologies is based on deep level defects in semiconductors, with the nitrogen-vacancy (NV) center in diamond being the most prominent example. Unfortunately, diamond in sufficient quality is scarce, which motivated efforts to find similar defects in silicon carbide (SiC). So far, many reports focus on investigations of point defects in irradiated 3C-SiC and as-grown material. However, the investigated defects are more or less a product of coincidence for both. While in irradiated material the intentional generation of specific defects is rather challenging, in as-purchased material the defects are actually more an unintentional by-product of growth and process conditions. This work proposes a new route: the incorporation and control of deep level defects in 3C-SiC by epitaxial sublimation growth. The observed defects in the near infrared show bright luminescence in the 175 K/200 K regime and remain excitable up to 300 K. This could enable working temperatures above the cryogenic limit. The joint origin of all detected defects is assigned to the carbon vacancy.

Silicon Carbide (SiC) is by now firmly established as material for power-electronics due to its outstanding and well-known physical properties[1–3]. Indeed, some of these advantageous characteristics like a wide band gap, high thermal conductivity, mechanical hardness and optical transparency also provide the optimum conditions for hosting stable and optically active defects in the crystal lattice. Therefore, SiC is gaining more and more interest as a material for quantum applications[4]. It has been demonstrated in the past that SiC can host a number of deep level defects suitable for spin-qubits and single-photon-sources (SPS)[5–16]. However, efforts to find defect based SPS have been made for some time. The most prominent example for addressable spin defects is the nitrogen vacancy (NV) center in diamond[17–19]. This defect is probably the best studied structure among all potential qubit candidates. Its ability to act as solid-state SPS up to room temperature offered for the first time the possibility to realize practicable systems for quantum computing and quantum key distribution (QKD). Excellent works have been published dealing with the requirements and potentials of both, diamond and SiC based quantum applications[4,20–22].

While diamond of high material quality is still not available in large quantities[23], SiC is commercially produced on wafer scale exhibiting a low defect density[24,25]. Therefore, SiC provides the clear advantage of being a mature material. Moreover, emission from deep levels in SiC lies in the near infrared (NIR) and often near the telecommunication window of optical fibers around 1.3 µm and 1.5 µm, therefore pointing towards quantum communication. In case of the cubic polytype of silicon carbide (3C-SiC), the advantages in comparison with hexagonal SiC and diamond are even larger. In hexagonal SiC, the stacking of Si-C bilayers gives rise to symmetry-inequivalent lattice sites of cubic or hexagonal character which results in a variety of electronic sub-levels lying close together. The higher symmetry of the cubic crystal lattice eliminates the problem of symmetry-inequivalent configurations and leads to more unique defect signatures[12]. This allows the individual addressability of defects which is the prerequisite for the application of e.g. integrated quantum optics. From a more practical perspective, 3C-SiC can be grown on silicon substrates by heteroepitaxy[26,27] and consequently allows full integration of 3C-SiC structures into silicon based devices. This opens up the way for direct interconnection of active and passive components such as SPS and integrated non-linear optics, respectively[28]. For applications in quantum telecommunication, deep level defects with emission in the NIR are most favorable. A variety of those defects such as the silicon vacancy ($V_{Si}$), the neutral divacancy ($V_C V_{Si}$) and the carbon vacancy carbon antisite pair ($V_C C_{Si}$) have already been investigated in SiC in the recent past[6,7,9,15,29,30].

Nevertheless, it needs to be mentioned that all defects investigated in previous works were created by one of the following routes: i) as byproducts during crystal growth or ii) by irradiation or rather implantation and subsequent annealing. In some cases the characterized material was designated as as-purchased or as-grown. However, from a crystal grower's point of view, terms like "as-purchased" or "as-grown" are very unspecific as such material could have gone through completely different processing steps during growth and post-treatment, depending on e.g. the material manufacturer's practices. The generation of point defects by irradiation with energetic particles like electrons, ions or protons allows finer control over placement and concentration of deep levels than for as-grown defects[4,29]. However, forecasting the type of defect produced by the damage is not trivial and requires e.g. numerical simulation. Moreover, post implantation annealing is usually necessary to form and activate specific structures. A proper adjustment of irradiation and annealing processes is essential. Hence, to a certain extent the generation of the desired deep levels is a product of coincidence.

To resolve the above stated open issues, a more from the bottom up approach involving crystal growth technology is presented. Here we show for the first time the in-situ generation and control of intentional point defects in bulk 3C-SiC during sublimation growth. Furthermore, it was demonstrated that in-situ doping can be used for defect-engineering and the creation of defect complexes from intrinsic and extrinsic defects. Besides already known vacancy-related defects, we report about clear indications for an aluminum-related defect that has not yet been generated or described elsewhere.

## Results

**Optical response of bulk 3C-SiC.** The use of sublimation growth for the preparation of 3C-SiC allows the incorporation of a variety of intrinsic and extrinsic defects during crystal growth. In contrast to defects created by irradiation or implantation and subsequent annealing, the evolution history of defects and therefore the final concentration within the 3C-SiC material is different. Nitrogen (N), boron (B) and aluminum (Al) doped freestanding, single crystalline 3C-SiC layers were prepared by epitaxial sublimation growth[31].

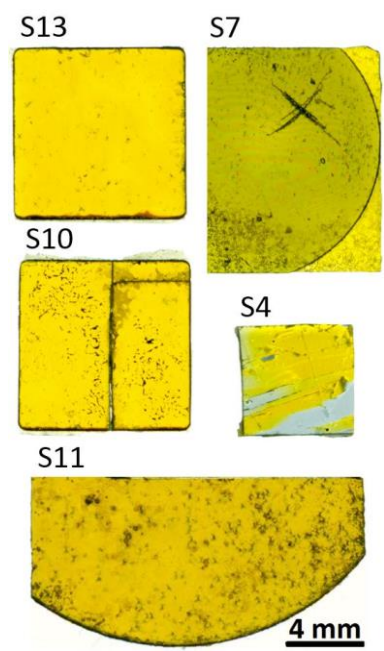

*Figure 1. Samples of bulk 3C-SiC prepared by epitaxial sublimation growth. Shown are samples S13, S7, S10, S4 and S11 as representatives for all samples prepared and characterized within this study. The dark areas in the samples originate from carbon inclusions, protrusions, cracks or surface inhomogeneities and were excluded from characterization. Grey areas in S4 originate from 6H-SiC polytype changes and were excluded from characterization as well.*

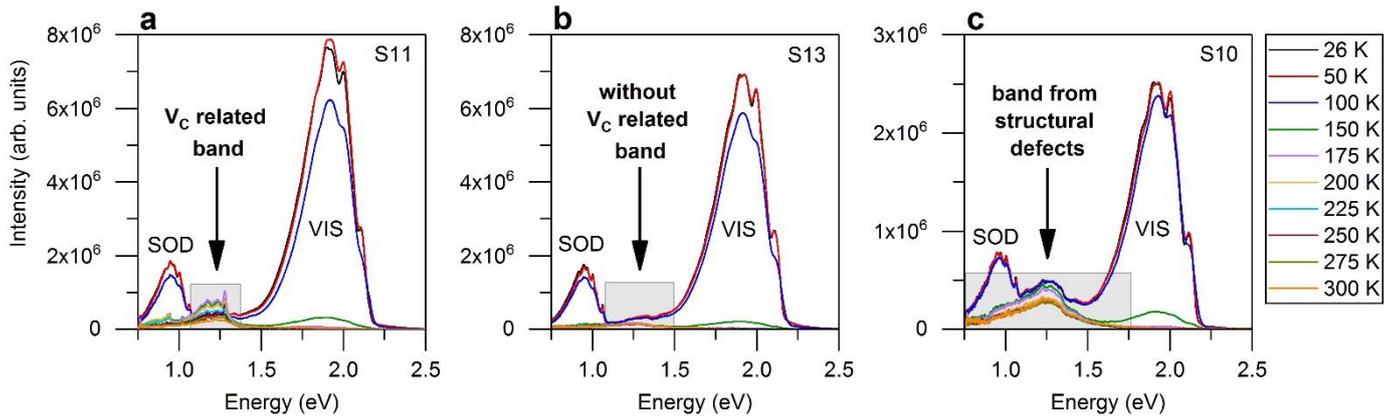

*Figure 2. Typical temperature dependent PL-spectra of sublimation grown 3C-SiC samples. The spectra were acquired with a 405 nm laser. Usually all spectra show a band in the visible range (VIS) and the luminescence of the related second order diffraction (SOD). Moreover, depending on the growth rate, luminescence in the NIR can be observed: (a) Spectrum of sample S11 which shows the NIR-band of $V_C$-related defects. (b) Spectrum of sample S13 which does not show the $V_C$-related NIR band. (c) Spectrum of sample S10 which shows a broad band of lower intensity. This band was only observed for a few samples grown at a low growth rate.*

In total, 15 samples referred to as S1-S15 were characterized within this work. Figure 1 shows a representative set of bulk 3C-SiC samples prepared and investigated within this study. Thickness and growth rate of the (100) and (111) oriented samples varied between 275 µm to 765 µm and 123 µm/h to 306 µm/h, respectively.

Raman and XRD-analysis verify that sublimation grown 3C-SiC is of high crystalline quality and quasi stress-free. Even though structural defects, especially stacking faults, still play a role, their density decreases with thickness and saturates below 1000 cm$^{-1}$ for layers thicker than 175 µm[32]. Temperature dependent photoluminescence measurements were performed between 26 K and 300 K with a step-size of 25 K using a closed-cycle He-cryostat. For above band gap excitation a 405 nm laser was used. Figure 2 depicts typical temperature dependent PL-spectra of sublimation grown 3C-SiC samples. Usually, all samples show a band in the visible range (VIS) and the related second order diffraction (SOD) of VIS. The intensity of the VIS-band and the related second order diffraction is high for low temperatures and decreases with increasing temperature. Samples grown within a limited growth rate regime (see trend in Figure 4) additionally exhibit a bright band consisting of sharp peaks within the near infrared (NIR). This band, shown in Figure 2a, increases and subsequently decreases with temperature reaching the maximum at 175 K or 200 K. It should be stated that a few samples grown at low growth rate exhibit another band that lies within the NIR-region (Figure 2c). This band differs considerably in shape and temperature-dependency of intensity from the usually observed NIR-band.

**Bright emission in the near infrared.** Within the NIR-band discrete peaks are apparent (see Figure 3). A weak indication of the NIR luminescence is already visible at 26 K. With increasing temperature, the NIR band increases as well until it reaches a maximum at 175 K or 200 K, due to the 25 K step size when performing PL analysis. With further increasing temperature, the NIR luminescence decreases again but remains excitable up to 300 K. Figure 3a illustrates the dependence of the NIR band from PL acquisition temperature. Spectra for temperatures below 150 K are not shown as they are influenced by luminescence from the VIS and SOD. In Figure 3b the maximum of NIR luminescence is shown for each sample. The diagram indicates that all samples within a limited growth-rate regime exhibit the same NIR-band with distinct peaks at the same positions. Altogether, four sharp peaks can be observed with center positions at 1.278 eV, 1.233 eV, 1.171 eV and 1.138 eV. Sample S7 is presented separately in Figure 3 due to an almost one order of magnitude higher intensity of the 1.171 eV peak.

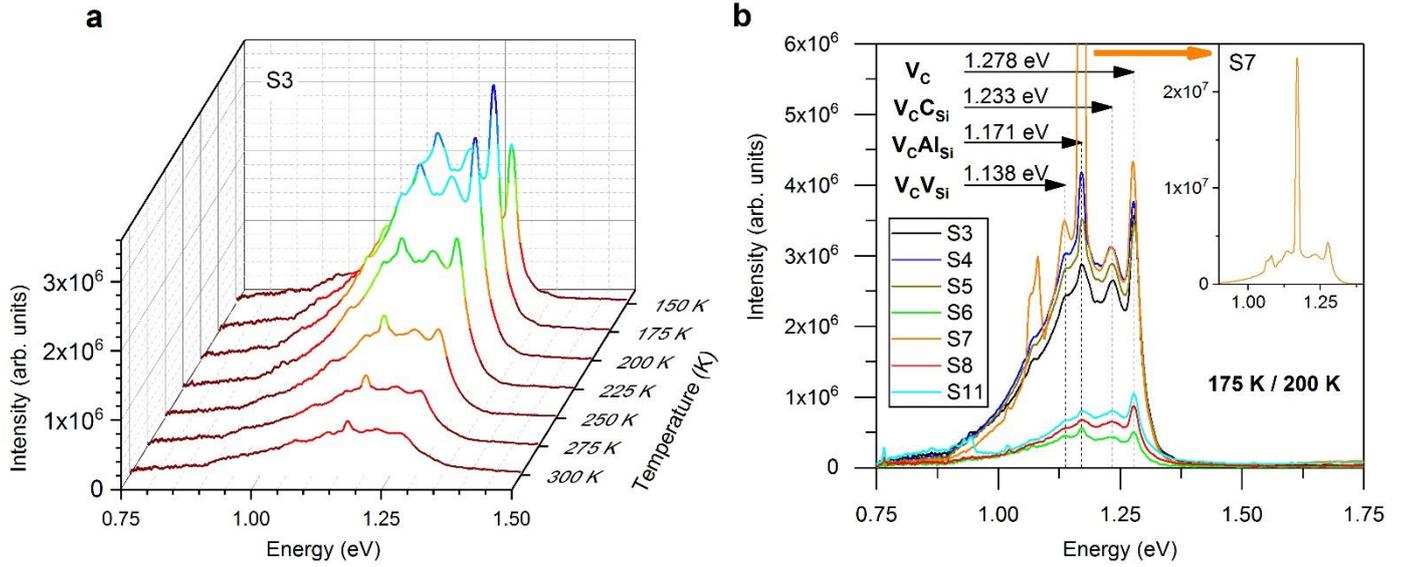

*Figure 3. Near infrared (NIR) luminescence of 3C-SiC: (a) NIR luminescence of sample S3 depending on PL acquisition temperature. (b) NIR luminescence of 3C-SiC samples prepared with different growth-rates by epitaxial sublimation growth. The diagram compares the spectra of maximum NIR intensity acquired at 175 K/200 K. Observed peaks and their assigned defects are indicated. The inset shows sample S7 with significant higher $V_CAl_{Si}$-intensity due to a higher relative Al-concentration. The dependence of NIR intensity on growth rate can be seen from Figure 4.*

**Growth-rate dependent generation of $V_C$-related defects.** Figure 4 shows the dependence of the $V_C$-peak intensity (a) and VIS-band intensity (b) on the growth-rate of 3C-SiC measured by photoluminescence at 175 K/200 K and 26 K, respectively. The intensity of the $V_C$-peak is zero for all samples with low growth-rates up to 179 µm/h (S2) (see Figure 4a). For samples with growth-rates between 203 µm/h (S8) and 270 µm/h (S4) the $V_C$-peak intensity increases before it starts to drop again for all samples with higher growth rates up to 306 µm/h (S6). Due to the step size (25 K) of PL-measurements the maximum NIR peak intensities are reached for 175 K or 200 K. In Figure 4b the intensity of the whole VIS-band at 26 K shows a stable luminescence for growth-rates up to approximately 243 µm/h (S3). For even higher growth rates a continuous decrease of luminescence over several orders of magnitude is observed.

**Detection of a novel Al-related defect.** The sharp peak at 1.171 eV shows a dependence (see Figure 5) on the relative aluminum concentration in 3C-SiC samples exhibiting the bright NIR band (see Figure 3). The results indicate a clear trend of increasing peak-intensity with increasing relative Al-concentration. Of course, the intensities of the presented samples are also affected by the growth rate and concentration quenching as can be seen from Figure 3 and Figure 4. However, taking sample S7 as an example, the considerable difference of the height of the 1.171 eV peak is clearly related to Al-concentration. These findings allow to assign this peak to an Al-related defect.

## Discussion

Conclusive results were obtained concerning luminescence from deep level point defects in 3C-SiC doped with different dopants and dopant concentrations. Not all samples showed the same luminescence behavior but all observed bands could be unambiguously identified. Two bands in the NIR and one band in the VIS were visible. In this work the luminescence in the NIR was investigated.

First of all, the results will be interpreted in relation to literature. Wang et al. [16] observed bright single photon emitters with center wavelengths ranging from 1080 nm to 1265 nm. The origin of the broad bands were assigned to stacking faults. This is in accordance with our results as we observed such broad and low intensity bands that might originate from structural defects, probably SFs, as well (see Figure 2c) for samples grown with lower growth rates. In our case these bands were not limited to a fixed narrow range in the spectra, but were found at different wavelengths within the near-infrared region for different samples.

Besides that, a second NIR band was observed (Figure 2a). This much more distinct band was of considerable higher intensity and showed four discrete peaks. This band was only present for 3C-SiC material produced with growth rates between 203 µm/h and 306 µm/h. According to literature, the discrete NIR luminescence can be assigned to various $V_C$-related defects. The center wavelength of these peaks were determined at positions of 1.138 eV, 1.171 eV, 1.233 eV and 1.278 eV. In literature there is still a controversy concerning intrinsic point defects and their defect complexes in 3C-SiC. Nevertheless, as a first step the available data from literature were used for the identification of the peaks.

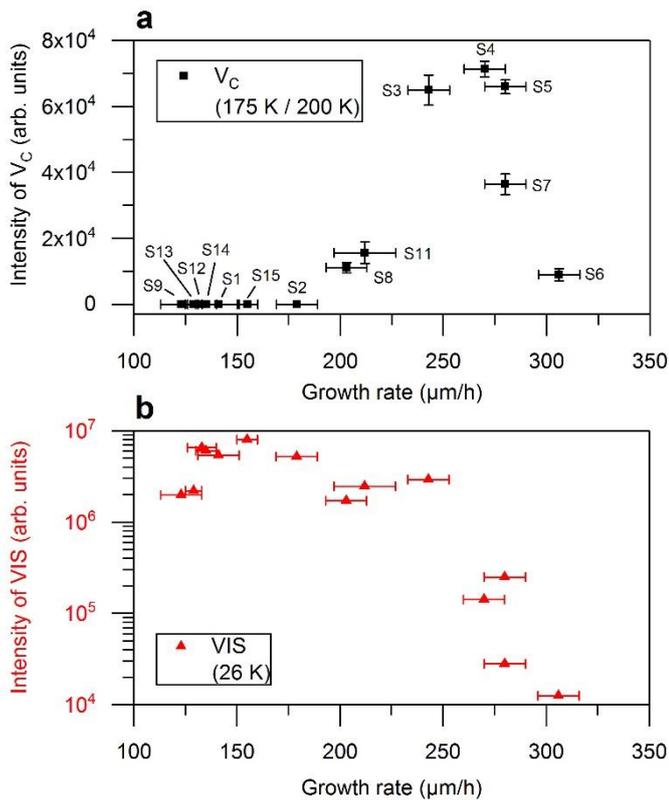

*Figure 4.* Intensity of the $V_C$-peak and the VIS-band luminescence as a function of growth rate: (a) Dependence of $V_C$-peak intensity on growth-rate measured by photoluminescence at 175 K/200 K. (b) Dependence of VIS-band intensity on growth-rate measured by photoluminescence at 26 K.

With that, the peak at 1.278 eV is assigned to the $V_C(+/++)$ defect, which is in good agreement with the value from literature of 1.29 eV[33]. The peak at a position of 1.233 eV is assigned to the carbon-vacancy carbon-antisite pair $V_CC_{Si}$ at 1.24 eV[33]. Castelletto et al.[9] reported about 3C-SiC nanoparticles hosting carbon-vacancy carbon-antisite pairs as single photon emitters. They observed broad peaks in the range of 650 nm. In contrast to our results in bulk material, the divergent results concerning position and broadness of the peaks lie in the nature of nanoparticles that are underlying spatial confinement effects. There are no reports in the literature that allow a clear assignment of the peak at 1.171 eV. As this peak is very sharp, it should not originate from structural defects. The luminescence at 1.138 eV can be assigned to the divacancy $V_CV_{Si}$, which is in very good agreement with the value of 1.13 eV[12] from literature.

So far, all results were discussed in the context of literature. In the following section the results will be explained from a crystal growth point of view. As can be seen from Figure 4a, we observed a growth rate dependent luminescence of $V_C$-related defects. In general, for vapor phase growth methods (PVT/SE), an increase of growth rate is, besides pressure, realized by increasing the temperature gradient and/or the absolute temperature at the growth interface, thus increasing the driving force for crystallization and therefore inducing a faster incorporation of the species from the gas phase into the growing crystal. The much faster transport enhances the general probability for incorporation of defects, especially vacancies. Epitaxial sublimation growth of 3C-SiC is moreover performed under Si-rich gas-phase conditions[34], hence offering much more Si gas-species than C gas-species at the growth interface. Even though this growth condition is a necessity to stabilize the cubic polytype during growth, it will further support the incorporation of carbon vacancies. Figure 4 shows the dependence of PL-intensity on growth-rate for both the NIR band at 175 K/200 K and the VIS band at 26 K. Due to the step-size of 25 K for the T-dependent PL-measurements, the actual intensity maxima for the NIR data points might differ slightly. Nevertheless, the trend for the NIR luminescence given in Figure **4**a is still reasonable. For low growth rates, no luminescence in the NIR can be observed. The following increase of luminescence is considered to be directly dependent on $V_C$-concentration.

Thereafter, the subsequent decrease of luminescence for even higher growth rates is attributed to concentration quenching[35–37] caused by $V_C$-related defects. This hypothesis is supported by the fact that the VIS band behaves in the same way for high growth rates. Whereas the intensity of the whole VIS band is stable for growth rates up to approximately 243 µm/h, the luminescence drops for all samples grown even faster. The evolution of the overall PL-intensity is a function of defect concentration. At low defect concentrations an increase of the number of defects causes an increase of PL yield. However, with increasing number of luminescent centers, the increase of defects causes a decrease of PL intensities. This is due to quenching effects such as non-radiative coupling of luminescent centers at high defect concentrations, also referred to as concentration quenching[35–37].

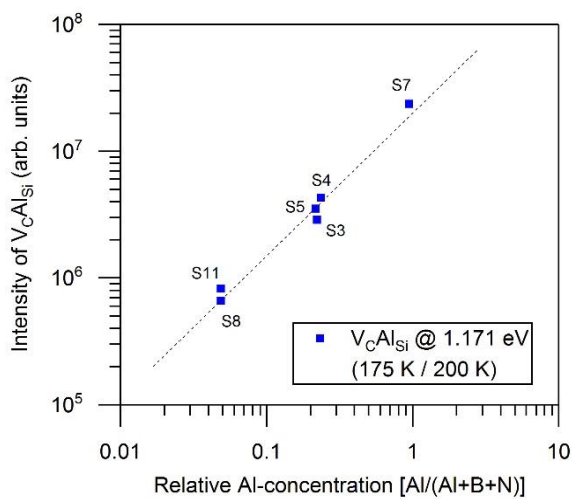

*Figure 5* Dependence of $V_CAl_{Si}$-peak intensity on relative aluminum concentration in the 3C-SiC sample.

The appearance of the broad band in the NIR (Figure 2c) should also be discussed in the context of crystal growth. Even if the origin of this band is not yet fully understood, there are indications hinting to structural defects[16]. However, the broad luminescence of lower intensity could also originate from interstitials, or both SF and interstitials[38–40]. At this point, the pioneering work of Voronkov[41] should be remembered who established the V/G-criterion, which describes the growth rate (V) and temperature gradient (G) dependent formation of interstitials and vacancies in silicon. A similar effect may be responsible for the growth rate dependent NIR emission in this work. This very broad band of lower intensity was only observed for a few samples grown at low growth rate. For low growth rates, the probability for the incorporation of interstitials is higher than for the generation of vacancies. Due to their structure and migration mechanisms, interstitials are more mobile within the crystal lattice than other point defects. In the case of high densities of structural defects, the interstitials can easily move towards structural defects which act as sinks for interstitials. It seems quite reasonable that such clusters of interstitials around structural defects could give rise to broad bands of low intensity as observed in this work.

In Figure 5 the intensity of the peak at 1.171 eV versus the relative Al-concentration is shown. Comparing the trend in Figure 5 with the peak intensities in Figure 3 it can be seen that the peak at 1.171 eV increases in intensity and in sharpness with increasing relative Al-concentration indicating a dependency from Al content. Note that sample S7 with the highest relative Al-concentration cannot be shown in the main Figure 3 but in the inset therein, due to a much higher intensity of the 1.171 eV peak. Hence, this finding is in accordance with the trend described above. All samples that show this intensive luminescence were grown within the regime of high $V_C$-concentration (see Figure 4). In the crystal lattice of 3C-SiC, Al preferably occupies the Si-site hinting to an Al-related defect, too[42]. In consideration of the general $V_C$-rich growth regime and the supposed occupation of the Si-site by Al, the peak observed is postulated to originate from the $V_CAl_{Si}$ defect. Of course this postulation needs to be verified in future works, e.g. by DFT simulations.

By variation of growth-rate, the NIR-luminescence of 3C-SiC can be specifically tailored for individual requirements. In order to get maximum luminescence intensity, the amount of optically active centers should be as large as possible but should not exceed the critical concentration where concentration quenching becomes relevant. For e.g. qubits, a low number of luminescent defect centers is necessary which can as well be achieved by using adequately low growth rates.

As the maximum NIR luminescence was observed in the temperature range between 175 K and 200 K, an application of such material may enable working temperatures above the cryogenic limit. This would allow for example simple cooling using dry ice.

For the $V_CV_{Si}$ and the $V_CC_{Si}$ it has already been demonstrated that these defects can behave as single photon sources on single defect level[4,29]. Even if further steps like isolation of single defects for SPS will be necessary to realize applications, $V_C$ related defects can be considered as highly promising candidates for SPS and future devices with quantum functionality.

In conclusion, we present the luminescence of various $V_C$-related defects in the near infrared region. The principal defect, as essential for the occurrence of all observed defects, is the carbon vacancy. Therefore, the whole NIR band reveals a growth rate dependence that can be very well explained by inherent growth conditions during epitaxial sublimation growth. For the first time, the controlled incorporation of deep level defects during sublimation growth of high-quality bulk 3C-SiC is presented. Additionally, doping opens up the possibility of defect engineering. By variation of the aluminum content, the intensity of the $V_CAl_{Si}$ defect could be changed by more than one order of magnitude.

Controlled incorporation of deep level defects in 3C-SiC, defect engineering through doping and the possibility of cryogenic cooling can bring practicable application of quantum devices within tangible reach.

## Methods

**Sublimation growth of 3C-SiC.** Nitrogen (N), boron (B) and aluminum (Al) doped freestanding, single crystalline 3C-SiC layers were prepared by epitaxial sublimation growth in our lab. Details of the growth setup are given elsewhere[31]. For the characterizations in this work, samples of 12x12 mm or pieces from 1'' layers were used. The characterization was performed on as-grown surfaces with negligible native surface-oxide. Dopant type and concentration was varied by the choice of the source material and determined by SIMS or by estimation, assuming similar dopant concentration of sample and source material. It should be noted that even though doping was varied from n-type to p-type by selecting N-, B- and Al doped source wafers of different dopant concentrations, all samples (S) are unintentionally co-doped to a certain extent. Thickness and growth rate of the (100) and (111) oriented samples varied between 275 µm to 765 µm and 123 µm/h to 306 µm/h, respectively.

**Photoluminescence characterization of 3C-SiC.** Temperature dependent photoluminescence measurements were performed between 26 K and 300 K with a step-size of 25 K using a closed-cycle He-cryostat. A laser diode with 405 nm (Coherent CUBE 405-100C) in combination with a 405 nm bandpass as well as a 450 nm long pass filter were used for above band gap excitation. All measurements were conducted with 50 mW laser power and a spot size of 900 µm in diameter. The penetration depths of the laser is approximately 10 µm in the case of high-quality 3C-SiC[43]. The spectra were acquired in the range between 450 nm and 1700 nm with a cooled InGaAs array detector (Horiba Symphony IGA-512 x 1) utilizing a monochromator Horiba TRIAX 552 (150 grating). All spectra were converted to energy-scale by applying Jacobian Conversion[44,45].

**Fitting procedure of photoluminescence spectra.** PL-Spectra were analyzed by fitting and approximating the zero-phonon-lines (ZPL) as well as phonon-side-bands (PSB) of the involved peaks in the NIR-band and integrating the whole VIS-band as a measure of total luminescence. It should be noted, that intensity was set to zero if a sample did not show the peak or band under investigation (see e.g. Figure 4). This was done to avoid falsification of the results due to background intensity from e.g. adjacent bands that exhibit as well a temperature dependence. All observed bands could be unambiguously identified and the resulting background intensities were significantly lower than the intensities under investigation.

## Data availability

The datasets generated or analyzed during the current study are available from the corresponding author (P.W.) on reasonable request.

**Acknowledgements**

The authors thank Ulrike Künecke for proofreading of the manuscript. Partial financial support by the European Union in the frame of the Horizon 2020 program (CHALLENGE project, grant number 720827) is greatly acknowledged.

**Author contributions**

M.S. and P.W. conceived and planned the experiments; M.S. and M.L. carried out the experiments and analyzed the data; M.S. wrote the manuscript with support from M.L. and P.W; P.S. supported the growth of samples; P.W. supervised the project; M.S., M.L. and P.W. discussed the results.

**Additional information**

**Competing interests:** The authors declare no competing interests.